\documentclass[showpacs,aps,pra,amsmath,amssymb,twocolumn]{revtex4}
\usepackage{graphicx}
\usepackage{dcolumn}
\usepackage{bm,color}
\usepackage{txfonts}
\usepackage{amsmath,epsfig}
\usepackage{epstopdf}
\newcommand{\ket}[1]{\left\vert#1\right\rangle}
\newcommand{\bra}[1]{\left\langle#1\right\vert}

\newcommand{\eq}{Eq.~}
\newcommand{\eqs}{Eqs.~}
\newcommand{\fig}{Fig.~}

\newcommand{\cf} {cf.~}
\newcommand{\ie} {i.e.~}

\begin{document}

\author{Francesco Ciccarello}
\affiliation{
CNISM and Dipartimento di Fisica, 
Universita' degli Studi di Palermo, Viale delle Scienze, Edificio 18, I-90128 Palermo, Italy}


\title{Resonant atom-field interaction in large-size coupled-cavity arrays}
 
\date{\today}
\begin{abstract}
We consider an array of coupled cavities with staggered inter-cavity couplings, where each cavity mode interacts with an atom. In contrast to large-size arrays with uniform-hopping rates where the atomic dynamics is known to be frozen in the strong-hopping regime, we show that resonant atom-field dynamics with significant energy exchange can occur in the case of staggered hopping rates even in the  thermodynamic limit. This effect arises from the joint emergence of an energy gap in the free photonic dispersion relation and a discrete frequency at the gap's center. The latter corresponds to a bound normal mode stemming solely from the finiteness of the array length. Depending on which cavity is excited, either the atomic dynamics is frozen or a Jaynes-Cummings-like energy exchange is triggered between the bound photonic mode and its atomic analogue. As these phenomena are effective with any number of cavities, they are prone to be experimentally observed even in small-size arrays.
\end{abstract}
\maketitle

\noindent

\section{Introduction}

The tremendous interest spread over the last few years in the dynamics of coupled-cavity arrays (CCAs) \cite{reviews} has made this emerging field a topical one within the general framework of quantum coherent phenomena and beyond. Many are the reasons behind such a substantial and widespread attention. Along with cold atoms in optical lattices \cite{sanpera} CCAs stand out as an attractive controllable test bed for many-body phenomena such as quantum phase transitions \cite{QPT}. An appealing feature of CCAs is \emph{local} addressing, namely the available high control of each single site in terms of performable measurements, quantum-state engineering and dynamical parameters tuning. From the perspective of quantum optics, CCAs open the door to the exploration of regimes where an almost ubiquitous feature of so-far-investigated atom-photon dynamics, \ie the effectiveness of descriptions in terms of single-atom dynamics, does not hold any more. 
Importantly, the variety of possible experimental setups prone to implement CCAs \cite{exp} and, mainly, the widespread expectation that the technology required for their actual fabrication is by now at hand (at least for small clusters) are providing formidable motivations to explore the rich physics of CCAs. In particular, one of the lines along which current investigations are proceeding is the study of excitation propagation in various forms such as transport of photons and excitons \cite{sun-photons, irish, greentree}, polaritons \cite{daniel, single-polariton} and solitons \cite{mauro}. In this framework, transport in CCAs with defects or impurity atoms is also receiving considerable attention \cite{longo, sun-quasi-bound, shi,nori-no-atomi}.

In a typical arrangement of CCAs, the field mode of each cavity is coupled to a two-level atom via a Jaynes-Cummings-type interaction. Energy can be thus stored in the form of both photons and excitons (atomic excitations) that are able to transform into each other as well as propagate along the array. While photons can move by direct hopping between neairest-neighbour cavities, excitons can travel only provided that they are transformed into photonic excitations and eventually converted back. The dynamics ruling excitation transport in CCAs is therefore non-trivial and physically attractive. While first works along this line addressed small clusters \cite{roversi,irish} only more recently arbitrary-size arrays have been tackled \cite{greentree}. In all such scenarios, for uniform values of atom-photon interaction strength, cavity-mode frequency and atomic detuning and whenever the decoupling of the field's hopping Hamiltonian in terms of normal modes is known the Hamiltonian describing the full many-body dynamics enjoys an attractive feature. It can indeed be rearranged \cite{irish,greentree} as the sum of decoupled Jaynes-Cummings (JC) models \cite{JC}, each coupling a field (bosonic) normal mode to its excitonic analogue. So long as the array size is small (assuming non-degenerate photonic normal modes) resonant excitation of only one of such effective JC systems is possible by judicious tuning of the atom frequency \cite{irish}. Physically, this circumstance is quite remarkable since it entails that significant amounts of energy can be exchanged between atoms and photons even in the strong-hopping regime, \ie when the atom-photon interaction rate is much lower than the photon hopping rate. Such a picture, however, can drastically change with large-size CCAs. By taking uniform-hopping rates, for instance, the passage to a large number of cavities brings about that the free field normal-mode spectrum tends to a continuous band. Clearly, this rules out the possibility of any resonance and thus first-order atom-field energy exchange. As a result, in the regime of strong hopping the atomic dynamics turns out to be frozen \cite{greentree}. One may wonder whether the above picture still holds for large-size arrays when a \emph{non}-uniform pattern of hopping strengths is considered. Motivated by some findings in the framework of spin chains \cite{cambridge} Makin \emph{et al.} \cite{greentree} investigated a parabolic distribution of hopping rates and found no significant changes in the strong-hopping regime compared to the uniform-hopping behavior. 

In the present work, we show that resonant atom-photon interaction can take place even under strong-hopping conditions in an arbitrary-size array with a \emph{staggered} pattern of hopping rates. In particular, we demonstrate that when certain array sites are initially excited and each atom has negligible detuning from the single-cavity field a JC-like dynamics involving a bound photonic normal mode interacting with its excitonic analogue is triggered. Among other effects, it entails a significant exchange of energy between field and atoms. This behavior basically stems from two features. First, the \emph{gapped} nature of the normal-frequency spectrum of the hopping Hamiltonian. Second, the occurrence under open boundary conditions (BCs) of a discrete frequency corresponding to a \emph{bound} normal mode at the center of the gap. Such circumstances make resonant atom-photon dynamics possible even in the thermodynamic limit.

The present paper is organized as follows. In Section \ref{H-sec}, we introduce the set-up which we focus on, an array with staggered hopping rates, and give the Hamiltonian that describes its dynamics. We then focus on the special case where the free field Hamiltonian reduces to that associated with a uniform-hopping array, which allows us to briefly review the dynamics of such arrays. In Section \ref{staggered}, we focus on the free field Hamiltonian in the general case, present exact solutions for its normal modes and associated frequencies and discuss its main properties. We then illustrate the regime which we focus on and give the corresponding effective representation of the full Hamiltonian. In Section \ref{dynamics}, we show how an initial atomic excitation propagates along the array and discuss the salient features of the excitation-transport dynamics. Finally, in Section \ref{conclusions}, we give some comments and draw our conclusions.

\section{Hamiltonian and review of the uniform-hopping setting} \label{H-sec}

We consider a finite-length array of $N$ low-loss cavities, where nearest-neighbour cavities are coupled together so as to allow for photon hopping. Each cavity sustains a single field mode of frequency $\omega_f$, which is  coupled at a rate $J$ to a two-level atom of Bohr frequency $\omega_a$ (here and throughout we use units such that $\hbar\!=\!1$). The Hamiltonian of the full system reads
\begin{equation}\label{H}
\hat{H}=\hat{H}_f+\hat{H}_a+\hat{H}_I\,\,,
\end{equation}
where 
\begin{eqnarray}
\hat{H}_f &=& \omega_f \sum^N_{x=1}
       \hat{a}_x^{\dagger}\hat{a}_x
 - \kappa \!\sum^{N-1}_{x=1}\, [1-(-1)^x \eta] 	\,
     \left( \hat{a}_{x+1}^{\dagger}\hat{a}_{x} +\rm{h.c.}\right),\,\,\label{H0f}\\
\hat{H}_a &=& 
\omega_a  \sum^N_{x=1}
      \hat{b}_x^{\dagger}b_x\,,\label{H0a}\\\
\hat{H}_{I} &=& 
J  \sum^N_{x=1}\left(
       \hat{b}_x \,\hat{a}_x^{\dagger}+  \hat{b}_x^{\dagger}\, \hat{a}_x \right)\label{HI}\,\,. 
      \end{eqnarray}
In \eqs (\ref{H0f})-(\ref{HI}), $\hat{a}_x$ ($\hat{a}^{\dagger}_{x}$) is a bosonic field operator that annihilates (creates) a photon at cavity $x$, whereas $\hat{b}_x$ ($\hat{b}^{\dagger}_{x}$) is an atomic operator that annihilates (creates) an exciton on the $x$th atom according to $\hat{b}_{x}\!=\!\left[\hat{b}^{\dagger}_{x}\right]^{\dagger}\!=\!\ket{g}_x\!\bra{e}$, where $\ket{g}_x$ ($\ket{e}_x$) is the ground (excited) state of the $x$th atom. 
$\hat{H}_f$ and $\hat{H}_a$ are the free Hamiltonians of the field and atoms, respectively, while $\hat{H}_I$ describes the atom-field coupling. 
Notice that atomic operators associated with different sites commute, \ie $[\hat{b}_x,\hat{b}_{x'}]\!=\![\hat{b}_x,\hat{b}^{\dagger}_{x'}]\!=\!0$ for any $x\!\neq\!x'$ since operators acting on different Hilbert spaces commute. On the other hand, it is easily checked that $\forall x$ $\hat{b}_x^2\!=\![\hat{b}_x^{\dagger}]^2\!=\!0$ and $\hat{b}_x\hat{b}_x^{\dagger}\!+\!\hat{b}_x^{\dagger}\hat{b}_x\!=\!1$.

The distinctive feature of Hamiltonian (\ref{H}) is the staggered pattern of hopping strengths. 
The odd (even) hopping rates, \ie those associated with the nearest-neighbour cavities 1-2, 3-4,... (2-3, 4-5,...), take the value $\kappa_1\!=\!(1\!+\!\eta)\kappa$ [$\kappa_2\!=\!(1\!-\!\eta)\kappa$)], where $\kappa$ is a hopping rate. Hence, \emph{two} different hopping strengths are periodically interspersed along the array. As by setting $\eta\!=\!0$ an array with uniform hopping rate $\kappa$ is retrieved  \cite{greentree} the dimensionless staggering parameter $\eta$ measures the distortion of such uniform-hoppings CCA. For $|\eta|\!=\!1$, the array reduces to a collection of independent two-cavity blocks (but one comprising a single cavity in the case of odd $N$). A sketch of the whole set-up is given in \fig1(a). The distorted tight-binding model \cite{peierls} specified by $\hat{H}_f$ has recently been harnessed (in its fermionic version) in Ref.~\cite{sun} for the sake of quantum state transfer. As we thoroughly discuss in the next section, regardless of the value taken by $\eta$, $\hat{H}_f$ in \eq(\ref{H0f}) can be exactly arranged in a diagonal form in terms of normal-mode field operators. In the remainder of this Section, though, we focus on the uniform-hopping case $\eta\!=\!0$.

For $\eta\!=\!0$, the free field  Hamiltonian can be expressed in terms of normal modes as \cite{cambridge,greentree}
\begin{equation} \label{H0f-diagonal}
\hat{H}_f\!=\!\sum_{k} \omega_{k}\, \hat{\alpha}^{\dagger}_{k}\hat{\alpha}_{k}\,\,,
\end{equation}
where
\begin{eqnarray}
k&=&\frac{2\pi m}{N+1}\,\,\,\,\,\,\,\,(m=1,...,N)\,\,,\\
\omega_k\!&=&\omega_f+2\kappa \cos{\frac{k}{2}}\label{omegak}\,\,,\\
\hat{\alpha}_k\!&=&\sqrt{\frac{2}{N+1}}\sum_{x=1}^N\,\!\sin{\left(\frac{k}{2} x\right)} \,\hat{a}_x\,\,.\label{alphak}
\end{eqnarray}
In \eqs(\ref{H0f-diagonal}) and (\ref{alphak}), $\hat{\alpha}_k$ and $\hat{\alpha}_k^{\dagger}$ are, respectively, bosonic annihilation and creation field operators associated with the $k$th photonic normal mode, which satisfy standard commutation rules. 

A feature of Hamiltonian (\ref{H}) is that atom-photon interaction strengths, cavity-mode and atomic frequencies are uniform throughout the cavity array. This allows to arrange it in an elegant and useful form \cite{irish,greentree} in terms of $N$ decoupled effective JC models according to 
\begin{equation}
\label{HJC}
\hat{H} = 
\sum_{k}
   \left[
     \omega_k \,\hat{\alpha}_k^{\dagger}\hat{\alpha}_k
      +\omega_a \, \hat{\beta}_{k}^{\dagger} \,\hat{\beta}_{k} +
      J \left(
         \hat{\alpha}^{\dagger}_k \hat{\beta}_{k}+ \hat{\beta}_k^{\dagger} \hat{\alpha}_{k} 
      \right)
   \right] \,\,,
\end{equation}
where the expansion of the atomic normal-mode operators $\hat{\beta}_k$'s in terms of site operators $\hat{b}_x$'s is apart from their different commutation rules fully analogous to \eq(\ref{alphak}). Notice that, unlike $\hat{\alpha}_k$'s, each $\hat{\beta}_k$ has the same  associated normal-mode frequency $\omega_a$.

In each effective JC-like Hamiltonian appearing in the decomposition (\ref{HJC}), a field normal mode couples to its atomic analogue through a JC-type interaction. As all of the atomic modes have the same frequency $\omega_a$ when $\omega_a\!=\!\omega_{k'}$ for a given $k'$ only the corresponding pair of field and atomic normal modes is, in principle, resonantly excited.
\begin{figure*}
 \includegraphics[width=0.55\textwidth]{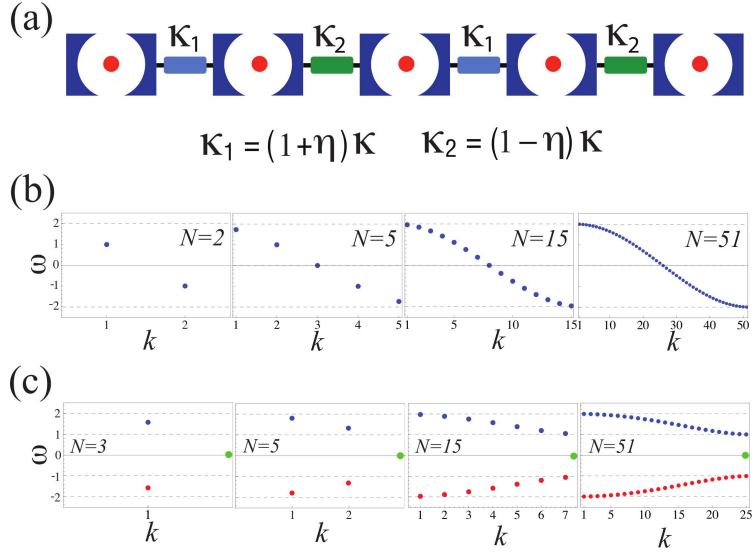}
\caption{(Color online) \textbf{(a)} Schematic sketch of an array of $N$ coupled cavities with staggered hopping strengths  $\kappa_1\!=\!(1\!+\!\eta)\kappa$ and $\kappa_2\!=\!(1\!-\!\eta)\kappa$. The field mode of each cavity is coupled to an atom (red dot). \textbf{(b)} $\eta\!=\!0$ (uniform-hopping case): normal-mode-frequency spectrum of the free field Hamiltonian $\hat{H}_f$  for $N\!=\!2,5,15,51$. Here and in next plot (\textbf{c}) frequencies are expressed in units of $\kappa$ and measured from $\omega\!=\!\omega_f$. (\textbf{c})  $\eta\!=\!0.5\,$: normal-mode-frequency spectrum of $\hat{H}_f$ using open BCs for $N\!=\!3,5,15,51$. The levels below (above) the $k$-axis [red (blue) dots] fall within the range $[\omega_f\!-\!2\kappa,\omega_f\!-\!\kappa]$ ($[\omega_f\!+\!\kappa,\omega_f\!+\!2\kappa]$), while a single discrete frequency $\omega\!=\!\omega_f$ (green dot) lies at the center of the gap between such intervals (its associated abscissa has been arbitrarily assigned).  \label{Fig1}}
\end{figure*}

For instance, for $N\!=\!2$ only two normal modes arise in the decomposition (\ref{H0f-diagonal}). Their associated annihilation operators and energies are in such a simple case conveniently expressed as \cite{irish} $\hat{\alpha}_{\pm}\!=\!(\hat{a}_1\!\pm\! \hat{a}_2)/\sqrt{2}$ and $\omega_{\pm}\!=\!\omega_f\!\pm\!\kappa$, respectively [see \fig1(b)]. Analogously, the atomic normal operators read $\hat{\beta}_{\pm}\!=\!(\hat{b}_1\!\pm\! \hat{b}_2)/\sqrt{2}$. When the atoms are tuned in resonance with one of these two modes by setting $\Delta\!=\!\pm \kappa$ (where $\Delta\!=\!\omega_f\!-\!\omega_a$ is the atoms' detuning from the single-cavity frequency) a JC-like dynamics occurs with a continuous exchange of energy between the involved photonic mode described by $\hat{\alpha}_{\pm}$ and the corresponding excitonic mode \cite{irish}. A typical behavior \cite{irish} that arises under such conditions is that when one atom, say atom 1, is initially prepared in the excited state (with atom 2 in the ground state and no photons populating the cavities) the photonic normal mode is progressively excited while the excitation probability of atom 2 grows. Over the following stage, while the field returns the received energy atom 2 keeps increasing its excitation probability until at a certain time the initial excitation has fully transferred to it (with the field and atom 1 unexcited). Afterward, the phenomenon is reversed through a further excitation-deexcitation cycle of the field mode until the initial state is fully retrieved. Triggering such a dynamics, which is of first-order in $J$, through a judicious tuning of the atomic frequency is however possible only in small-size arrays. 

Indeed, as $N$ grows [see \fig1(b)] more and more photonic levels gather in the range $[\omega_f-2\kappa,\omega_f+2\kappa]$ as implied by \eq(\ref{omegak}). In the thermodynamic limit $N\!\rightarrow\!\infty$, the photonic spectrum takes the form of a continuos band. For large arrays this in fact suppresses any possible resonant atom-photon coupling due to the lack of discrete frequencies. As a result, in the strong-hopping regime $J\!\ll\!\kappa$ a straightforward use of the interaction picture shows that so long as second-order processes are negligible the atomic dynamics becomes fully frozen \cite{greentree}. Hence, in striking contrast to the above-described process in a two-cavity array, an exciton initially localized on a given atom is unable to move away. Also, energy exchange between atoms and field is suppressed.

\section{Normal modes for staggered hopping rates and effective Hamiltonian} \label{staggered}

The scenario discussed above for a uniform-hopping array substantially changes when $\eta\!\neq\!0$. To illustrate this, in \fig1(c) we have numerically computed the normal-mode frequencies of the free field Hamiltonian (\ref{H0f}) for $\eta\!=\!0.5$ and $N\!=\!3,5,15,51$. Two major differences appear compared to the case $\eta\!=\!0$. First, the spectrum is gapped and thus in the limit of large $N$ two continuous bands emerge, instead of a single one. Second, regardless of $N$ a discrete normal frequency lies exactly at the center of the gap. These findings can be made analytically rigorous. 

To begin with, for the sake of notation compactness we define the two quantities
\begin{eqnarray}
\tau&=&\frac{\eta+1}{\eta-1}\,\,,\label{tau}\\
\varepsilon_k&=&2\kappa \sqrt{\cos^2 \frac{k}{2}\!+\!\eta^2\sin^2{\frac{k}{2}}}\label{epsk}\,\,.
\end{eqnarray}
Note that $\varepsilon_k$ is defined positive.

It can be checked that for an arbitrary odd number of cavities $N$ (later on we comment on this assumption) $\hat{H}_f$ can be exactly arranged in the diagonal form 
\begin{equation} \label{Hf-diag}
\hat{H}_f=\omega_{f}\, \hat{\alpha}_{\ell}^{\dagger}\hat{\alpha}_{\ell}+\sum_k \sum_{\mu=\pm} \omega_{k,\mu}\, \hat{\alpha}_{k,\mu}^{\dagger}\!\hat{\alpha}_{k,\mu}\,\,,
\end{equation}
where $k\!=\!2\pi m/(N\!+\!1)$ with $m\!=\!1,2,...,(N\!-\!1)/2$ and
\begin{eqnarray}
\hat{\alpha}_{\ell}\!&=&\!\frac{2}{\eta\!-\!1}\sqrt{\frac{\eta}{\tau^{N\!+\!1}\!-\!1}}\,\sum_{x=1}^{\frac{N+1}{2}}\,\tau^{x-1}\,\hat{a}_{2x-1}\,\,,\label{local}\\
\omega_{k\,\pm}\!&=&\omega_f \mp\,\varepsilon_k\label{energies}\,\,,\\
\hat{\alpha}_{k\,\pm}\!&=&\!\!\sqrt{\frac{2}{N\!+\!1}} \left( \sum_{x=1}^{\frac{N-1}{2}}\,\sin{(k x)} \,\hat{a}_{2 x} \pm\! \sum_{x=1}^{\frac{N+1}{2}}\! \sin{(k x\!+\!\vartheta_k)}\,\hat{a}_{2x-1}	\right)\,\,.\,\,\,\,\,\,\,\,\,\,\,\,\,\label{states}
\end{eqnarray} 
The phase $\vartheta_k$ appearing in \eq(\ref{states}) obeys
\begin{equation} \label{tetak}
e^{i \vartheta_k}=\frac{\kappa\,(1\!-\!\eta)}{\varepsilon_k}\,\left(e^{-i k}-\tau\right)\,\,.
\end{equation}
The discrete spatial functions specifying the expansion of normal operators \eqs(\ref{local}) and (\ref{states}) in terms of site operators $\{\hat{a}_x\}$ can be shown to fulfill orthonormality conditions, a proof that we carry out in Appendix \ref{AppB}. As such, $\{\hat{\alpha}_{\ell},\hat{\alpha}_{k,\pm}\}$ form a set of bosonic annihilation operators fulfilling standard bosonic commutation rules. As is to be expected, the normal-frequency spectrum $\{\omega_f,\omega_{k\pm}\}$ and corresponding set of normal annihilation operators $\{\hat{\alpha}_\ell,\hat{\alpha}_{k\pm}\}$ reduce to (\ref{omegak}) and (\ref{alphak}), respectively, in the special case $\eta\!=\!0$. This check is carried out in detail in Appendix C.

At variance with the problem tackled in Ref.~\cite{sun} that allowed assumption of cyclic BCs, in deriving \eqs(\ref{Hf-diag})-(\ref{energies}) we have used open BCs, which are the natural ones to impose in a finite-length CCAs \cite{greentree}. Similarly to the uniform-hopping case \cite{cambridge}, the solutions in the continuous spectrum [see \eq(\ref{states})] are straightforwardly obtained by superposing those fulfilling cyclic BCs in the case of $N\!+\!1$ sites \cite{sun} and then requiring that they obey hard-wall BCs. The discrete solution $\hat{\alpha}_{\ell}$ is readily found by direct demonstration once one imposes that its normal frequency be $\omega_f$ (so that the hopping term in $\hat{H}_f$ vanishes). 
In Appendix \ref{AppD}, we show in detail that upon replacement of (\ref{energies}), (\ref{local}) and (\ref{states}) on the right-hand side of \eq(\ref{Hf-diag}) the free field Hamiltonian in \eq(\ref{H0f}) is  retrieved.

The two aforementioned bands in the normal-frequency spectrum are analytically described by \eqs(\ref{epsk}) and (\ref{energies}) [see \fig1(c)] with their associated annihilation operators given by \eq(\ref{states}). Their energy gap $\Delta \omega$, which coincides with the one obtained with cyclic BCs \cite{sun}, fulfills 
\begin{equation} \label{gap}
\Delta\omega\ge4 \kappa |\eta|\,\,,
\end{equation}
where the identity occurs in the thermodynamic limit $N\!\rightarrow\!\infty$.

For $\eta\!\rightarrow\!0$ the gap width vanishes so as to retrieve the uniform-hopping case \cite{irish,greentree} analyzed in the previous section. As $|\eta|$ grows the lower bound of the gap on the right-hand side of (\ref{gap}) linearly increases at a rate proportional to $\kappa$. 
 \begin{figure}
 \includegraphics[width=0.45\textwidth]{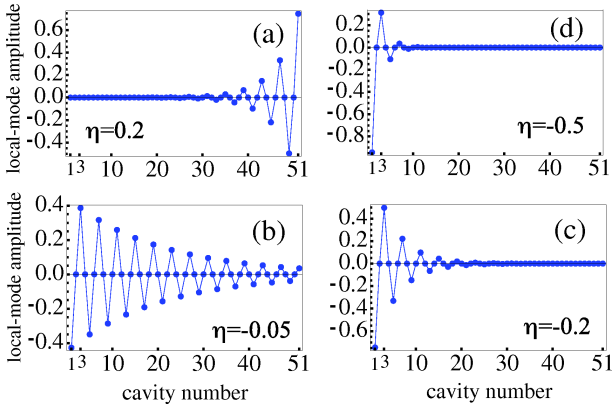}
\caption{(Color online) Amplitude of the bound mode in \eq (\ref{local}) against the cavity number for $N\!=\!51$ and $\eta\!=\!0.2$ (\textbf{a}), $\eta\!=\!-0.05$ (\textbf{b}), $\eta\!=\!-0.2$ (\textbf{c}) and $\eta\!=\!-0.5$ (\textbf{d}). As the array length $N$ only affects the normalization factor, when $N$ is varied the shape of these curves is unaffected. Notice that the amplitude vanishes at even sites. \label{Fig3}}
\end{figure}
As for the normal mode in \eq(\ref{local}), its associated frequency $\omega_f$ lies out of the two bands. Indeed, such normal mode is \emph{bound} in that, differently from modes (\ref{states}), it does not extend over the entire array. This is evident from the exponential functional form in \eq(\ref{local}) and \fig2, which shows that the mode amplitude decreases (increases) exponentially with the cavity number $x$ for negative (positive) values of $\eta$. Notice that while the mode amplitude strongly depends on $\eta$ its frequency is fully independent of both $\eta$ and $\kappa$. Using \eqs(\ref{tau}) and (\ref{local}), the characteristic length $\lambda$ over which the bound mode is spread is found as $\lambda\!=\!|1/\log[(1 \!+\! \eta)/(1\!-\! \eta)]|$. Thus the local-mode length becomes infinite when $\eta\!\simeq\!0$, as expected, while in the range $0\!<|\eta|\!<\!1$ it progressively shrinks for growing $|\eta|$ until it becomes negligible when $|\eta|\!\simeq\!1$.
This behavior is highlighted in \fig2, where the mode's spatial profile for different $\eta$'s is reported. A further feature which is evident in \eq(\ref{local}) and \fig2 is that regardless of $\eta$ the bound mode vanishes at even sites.

We now turn our attention to the full Hamiltonian (\ref{H}) in order to highlight the implications of \eqs(\ref{Hf-diag})-(\ref{states}) on the CCA's dynamics. We first define atomic normal operators in full analogy with \eqs(\ref{local}) and (\ref{states}) as 
\begin{eqnarray}
\hat{\beta}_{\ell}\!&=&\!\frac{2}{\eta\!-\!1}\sqrt{\frac{\eta}{\tau^{N\!+\!1}\!-\!1}}\,\sum_{x=1}^{\frac{N+1}{2}}\,\tau^{x-1}\,\hat{b}_{2x-1}\,\,,\label{localb}\\
\hat{\beta}_{k\,\pm}\!&=&\!\!\sqrt{\frac{2}{N\!+\!1}} \left( \sum_{x=1}^{\frac{N-1}{2}}\,\sin{(k x)} \,\hat{b}_{2 x} \pm\! \sum_{x=1}^{\frac{N+1}{2}}\! \sin{(k x\!+\!\vartheta_k)}\,\hat{b}_{2x-1}	\right)\,\,,\,\,\,\,\,\,\,\,\,\,\,\,\,\label{statesb}
\end{eqnarray} 
where again $k\!=\!2\pi m/(N\!+\!1)$ with $m\!=\!1,2,...,(N\!-\!1)/2$ with $\vartheta_k$ and $\tau$ still given by \eqs(\ref{tetak}) and (\ref{tau}), respectively. It is worth pointing out that unlike $\{\hat{\alpha}_\ell,\hat{\alpha}_{k\mu}\}$ the set $\{\hat{\beta}_\ell,\hat{\beta}_{k\mu}\}$ do not obey commutation rules since, while atomic site operators associated with different cavities commute, $\hat{b}_x\hat{b}_{x}^\dagger\!=\!1\!-\!\hat{b}_x^\dagger\hat{b}_{x}$ for any $x$.
In analogy with the uniform-hopping case in the previous Section, the full Hamiltonian in terms of field and atomic operators $\{\hat{\alpha}_\ell,\hat{\alpha}_{k\pm},\hat{\beta}_\ell,\hat{\beta}_{k\pm}\}$ takes the form
\begin{eqnarray}\label{HNMD}
\hat{H}&\!=\!&\hat{H}_{\ell}\!+\!\sum_{k\,\mu\!=\!\pm}\!\left[\omega_{k\pm}\hat{\alpha}_{k\mu}^{\dagger}\hat{\alpha}_{\kappa\mu}\!+\!\omega_a\,\hat{\beta}_{k\mu}^{\dagger}\,\hat{\beta}_{k\mu}\!+\!J(\hat{\beta}_{k\mu}^{\dagger} \hat{\alpha}_{k\mu}\!+\!{\rm h.c.})\!\right]\,\,,\,\,\,\,\,\,\,\,\,\,\,\,
\end{eqnarray}
where
\begin{equation} \label{Hell}
\hat{H}_{\ell}\!=\!\omega_f\hat{\alpha}_\ell^{\dagger}{\alpha}_\ell+\omega_a\,\hat{\beta}_\ell^{\dagger}\,{\beta}_\ell+J\left(\hat{\beta}_{\ell}^{\dagger} \hat{\alpha}_{\ell}\!+\!\hat{\alpha}_\ell\,\hat{\beta}_\ell^{\dagger}\right)\,\,.
\end{equation}
As we show in Appendix \ref{AppE}, the full Hamiltonian normal-mode decomposition (\ref{HNMD}) straightforwardly follows from decomposition (\ref{Hf-diag}) and the fact that $\omega_a$ and $J$ do not depend on the cavity site.

We take as free Hamiltonian $\hat{H}_0\!=\!\hat{H}_f\!+\!\hat{H}_a$ [see \eqs (\ref{H0f}) and (\ref{H0a})].
Due to $[\hat{b}_x,\hat{b}_{x'}]\!=\!0$ for any $x\!\neq\! x'$, the commutator between each site atomic operator and the free Hamiltonian is given by $[\hat{b}_x,\hat{H}_0]\!=\!\omega_a\hat{b}_x$, which immediately yields $[\hat{\beta}_\xi,\hat{H}_0]\!=\!\omega_a\,\hat{\beta}_\xi$ ($\xi\!=\!\ell,\{k,\mu\}$). Hence, in the interaction picture each normal atomic operator evolves with time according to $\hat{\beta}_\xi^{(I)}(t)\!=\!\hat{\beta}_{\xi}e^{-i \omega_a t}$ ($\xi\!=\!\ell,\{k,\mu\}$). On the other hand, using \eq(\ref {Hf-diag}) the normal field operators evolve in the same picture as $\hat{\alpha}_\ell^{(I)}(t)\!=\!\hat{\alpha}_{\ell}e^{-i \omega_f t}$ and $\hat{\alpha}_{k\mu}^{(I)}(t)\!=\!\hat{\alpha}_{k\mu}e^{-i \omega_{k\mu} t}$. In the interaction picture, the interaction Hamiltonian [\cf \eq(\ref{HNMD})] thus reads
\begin{equation}\label{HII}
\hat{H}_{I} ^{(I)}\!=\!J\hat{\alpha}_{\ell}^{\dagger}\, \hat{\beta}_{\ell}\,e^{i \Delta t}+J\!\sum_{k\mu=\pm}\!
          \left(\hat{\alpha}_{k\mu}^{\dagger} \,\hat{\beta}_{k\mu}\,e^{i (\omega_{k,\mu}-\omega_a) t} 
    \right)+
{\rm h.c.}\,\,
      \end{equation}
When $\Delta\!=\!0$, \ie each atom is on resonance with its own cavity mode, \eq(\ref{HII}) shows that while the contribution to $\hat{H}_{I}$ from the bound mode is constant in time, those arising from the normal modes in the two energy bands rotate at frequencies that are at least equal to half the energy-gap lower bound $2\kappa|\eta|$ [see (\ref{gap})]. Hence, provided that
\begin{equation}\label{regime}
J/\kappa\ll|\eta|\,\,,
\end{equation}
such contributions are rapidly rotating and thus do not affect the system's dynamics. When $\Delta\!\neq\!0$, this conclusion still holds provided that the detuning is at most of the same order of magnitude of $J$ and obeys $|\Delta|\!\ll\!\kappa|\eta|$. Therefore, in the above regime the effective Hamiltonian in the Schr\"odinger picture reduces to 
\begin{eqnarray}\label{Heff}
\hat{H}_{\rm eff}&\!=\!&\hat{H}_{\ell}\!+\!\sum_{k\,\mu\!=\!\pm}\!\left(\omega_{k\pm}\hat{\alpha}_{k\mu}^{\dagger}\hat{\alpha}_{\kappa\mu}\!+\!\omega_a\,\hat{\beta}_{k\mu}^{\dagger}\,\hat{\beta}_{k\mu}\right)\,\,.\,\,\,\,\,\,\,\,\,\,\,\,
\end{eqnarray}

\eq (\ref{Heff}) embodies a central finding of this work, namely the possibility that the complex many-body atom-photon interaction reduces to an effective JC-like coupling described by $\hat{H}_{\ell}$ between a bound photon mode and its excitonic analogue with all of the remaining field and atoms' modes freely evolving. According to \eq(\ref{regime}), when $|\eta|\!\le\!1$ such as in the numerical examples in \fig2 the occurrence of this few-degree-of-freedom dynamics is fully compatible with the strong-hopping regime $J\!\ll\!\kappa$. In the thermodynamic limit, \ie in practice for large-size CCAs, this marks a major difference from the uniform-hopping array \cite{greentree} discussed in the previous section.

\section{Dynamics of excitation transport} \label{dynamics}

Our next goal is to shed light on the features of the system's time evolution in the one-excitation subspace in line with Refs.~\cite{irish,greentree}. 
We recall that, despite normal atomic operators associated with different modes in general do not commute, in the one-excitation subspace they effectively do \cite{nota-comm}.

We denote by $\ket{0}$ the system's state with zero excitations, either photonic or atomic, and define states  $\ket{\Psi_{\pm}}$ as
\begin{equation}\label{Psi}
\ket{\Psi_{\pm}}=A_{\pm}\,\hat{\alpha}_\ell^{\dagger}\ket{0}+B_{\pm}\,\hat{\beta}_\ell^{\dagger}\ket{0}\,\,,
\end{equation}
with
\begin{eqnarray}
A_{\pm}=\frac{2 J}{\sqrt{(\Delta\pm\Omega)^2+4J^2}}\,\,,\\
B_{\pm}=\frac{\Delta\pm\Omega}{\sqrt{(\Delta\pm \Omega)^2+4J^2}}\,\,. 
\end{eqnarray}
Here, $\Omega\!=\!\sqrt{\Delta^2+4J^2}$ is the usual Rabi frequency associated with the JC-like Hamiltonian (\ref{Hell}). $\ket{\Psi_{\pm}}$ are eigenstates of $\hat{H}_{\ell}$, and hence of $\hat{H}_{eff}$, with eigenvalues $(\omega_a\!+\!\omega_f)/2\pm\Omega/2$. Using this and taking as paradigmatic initial state $\ket{\Psi(0)}$ one such that a given atom $x_0$ is excited, \ie $\ket{\Psi(0)}\!=\!\hat{b}^{\dagger}_{x_0}\ket{0}$, at a later time $t$ the system has evolved according to
\begin{eqnarray}\label{decomposition}
\ket{\Psi(t)}\!&=&\!\bra{0}\hat{\beta}_\ell\,\hat{b}_{x_0}^{\dagger}\ket{0}\left(B_+ e^{-i \Omega/2 t}\ket{\Psi_+}\!+\!B_- e^{i \Omega/2 t}\ket{\Psi_-}\right)\nonumber\\
&\!+\!&\sum_{k}\!\sum_{\mu=\pm}\bra{0}\hat{\beta}_{k,\mu}\,\hat{b}_{x_0}^{\dagger}\ket{0}\,e^{i \Delta/2 t}\,\beta_{k,\mu}^{\dagger}\ket{0}\,,\,\,\,\,\,\,\,\,\,\,\,\,
\end{eqnarray}
up to an irrelevant phase factor. \eq(\ref{decomposition}) was obtained by expanding $\ket{\Psi(0)}$ in the basis of stationary states of $\hat{H}_{eff}$ $\{\ket{\Psi_{\pm}},\hat{\alpha}_{k,\pm} \ket{0}, \hat{\beta}_{k,\pm}\ket{0}\}$. In the following analysis of the implications of \eq(\ref{decomposition}), we shall make use of the completeness of the basis of single-photon states $\{\hat{\alpha}_\ell^{\dagger}\ket{0},\hat{\alpha}_{k,\pm}^{\dagger}\ket{0}\}$, \ie
\begin{equation}\label{completeness}
\hat{\alpha}_\ell^{\dagger}\ket{0}\!\bra{0}\hat{\alpha}_\ell+\sum_{k}\sum_{\mu=\pm}\hat{\alpha}_{k,\mu}^{\dagger}\ket{0}\bra{0}\hat{\alpha}_{k,\mu}=\openone_{1\rm{ph}}\,\,,
\end{equation}
where $\openone_{1\rm{ph}}$ is the identity operator in the one-photon Hilbert space of the field. \eq(\ref{completeness}) straightforwardly follows from orthonormality identities (\ref{ort1})-(\ref{ort3}).

Let us first consider the case that $x_0$ is even. As the bound mode (\ref{local}) does not overlap even sites (see previous section) we trivially have  $\bra{0}\hat{\beta}_\ell\,\hat{b}_{x_0}^{\dagger}\ket{0}\!=\!0$ so that only terms proportional to $\hat{\beta}_{k,\mu}^{\dagger}\ket{0}$ do contribute to $\ket{\Psi(t)}$. This immediately yields that $\bra{0}\hat{a}_x\ket{\Psi(t)}\!=\!0$ for any $x$, \ie no field excitation is developed. As for $\bra{0}\hat{b}_x\ket{\Psi(t)}$, namely the probability amplitude to find the $x$th atom excited, use of \eq(\ref{completeness}), which is clearly valid for operators $\hat{\beta}_{k\pm}$'s as well, along with \eq(\ref{states}) entails $\bra{0}\hat{b}_x\ket{\Psi(t)}\!=\!\delta_{x,x_0}$. In other words, when $x_0$ is even $\ket{\Psi(t)}\!=\!\ket{\Psi(0)}$, \ie the atomic excitation is frozen analogously to the uniform-hopping case \cite{greentree}.  The freezing behavior however may not occur when $x_0$ is \emph{odd}. Indeed, through a reasoning similar to the one carried out above it is immediate to prove that for odd $x_0$ if $x$ is even then both $\bra{0}\hat{a}_x\ket{\Psi(t)}$ and $\bra{0}\hat{b}_x\ket{\Psi(t)}$ vanish, namely the initial excitation can only spread over odd cavities. On the other hand, when $x$ is odd projection of \eq(\ref{decomposition}) onto $\hat{a}_x^{\dagger}\ket{0}$ and  $\hat{b}_x^{\dagger}\ket{0}$ respectively yield
\begin{eqnarray}\label{projectiona}
\!\bra{0}\hat{a}_x\ket{\Psi(t)}\!&=&-2i\,\mathcal{N}\,\,\!\tau^{\frac{x_0+x\!-\!2}{2}}\!\frac{J}{\Omega}\,\sin{\frac{\Omega}{2}t}\,\,e^{-i\frac{\Delta}{2}t}\,,\,\,\,\,\,\,\,\,\,\,\,\,
\\
\!\bra{0}\hat{b}_x\ket{\Psi(t)}\!&=&\!\delta_{x,x_0}\!+\!\mathcal{N}\tau^{\frac{x_0\!+\!x\!-\!2}{2}}\!\left[\left(\cos\!{\frac{\Omega}{2}t}\!-\!i\frac{\Delta}{\Omega}\sin\!{\frac{\Omega}{2}t}\right)e^{-i\frac{\Delta}{2} t}\!-\!1\right]\!,\,\,\,\,\,\,\,\,\,\,\label{projectionb}
\end{eqnarray}
where $\mathcal{N}$ is the square of the factor out of the sum in \eq(\ref{local}).
 \begin{figure*}
\bigskip
 \includegraphics[width=0.85\textwidth]{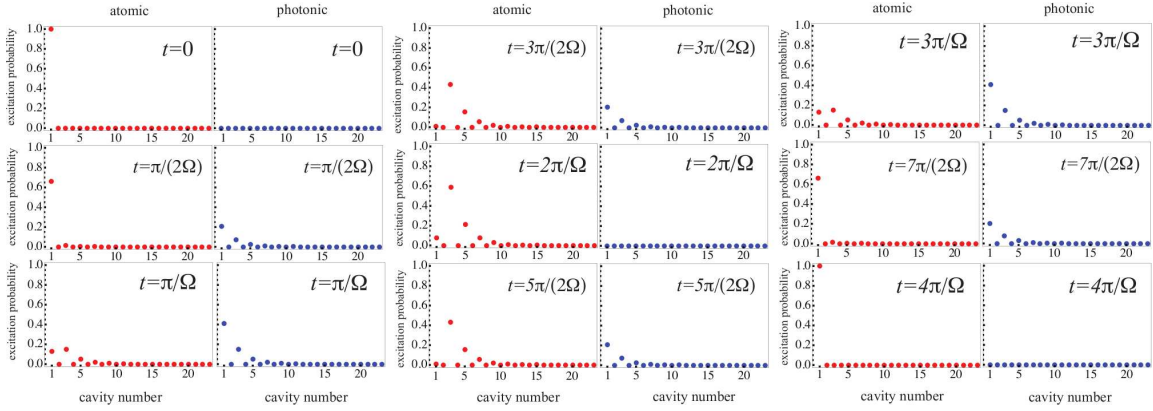}
\caption{(Color online) Snapshots of the atomic (red dots) and photonic (blue dots) excitation probabilities against the cavity number at time instants within the range $t\!\in\![0,4\pi/\Omega]$ and for the initial state where atom 1 is fully excited. We have set $\eta\!=\!-0.25$, $N\!=\!101$ and (in units of $J$) $\kappa\!=\!100$, $\omega_f\!=\!1000$, $\Delta\!=\!0$.  Although the plots were obtained from numerical solutions of the exact Hamiltonian $\hat{H}$, in practice they remain identical when \eqs (\ref{projectiona}) and (\ref{projectionb}) are employed. Only cavity numbers such that both excitation probabilities significantly differ from zero are shown. \label{Fig4}}
\end{figure*}
\eqs (\ref{projectiona}) and (\ref{projectionb}) fully describe the time evolution of the array in the regime such that the system Hamiltonian is well approximated by $\hat{H}_{eff}$ as given in \eq (\ref{Heff}). To illustrate the essential features of the dynamics, in \fig3 we address the case of $x_0\!=\!1$  and show several snapshots of the atomic and photonic excitation probalities, \ie respectively $|\!\bra{0}\hat{b}_x\ket{\Psi(t)}\!|^2$ and $|\!\bra{0}\hat{a}_x\ket{\Psi(t)}\!|^2$, in the time interval $t\!\in\![0,4\pi/\Omega]$ (afterwards the same behavior is  cyclically re-exhibited). The exciton initially present at cavity 1 (so that  at $t\!=\!0$ only the atomic bound mode is excited) progressively spreads over nearby odd cavities in the form of both field and atomic excitations. Such a stage, during which the field bound mode absorbs energy from the atomic one, finishes at time $t\!=\!\pi/\Omega$ when the field bound mode attains its maximum amplitude. Next, the field mode returns energy while the exciton probability spreading continues. At $t\!=\!2\pi/\Omega$, the field is again fully unexcited but, remarkably, the exciton is no more localized on the first atom. Rather, for each atom in an odd cavity (except cavity 1) the excitation probability reaches its maximum value. Finally, between $t\!=\!2\pi/\Omega$ and $t\!=\!4\pi/\Omega$ while the field undergoes a further excitation-deexcitation cycle analogous to the previous one  the excitonic distribution progressively localize around the first cavity until at $t\!=\!4\pi/\Omega$ the initial state is fully retrieved. All of such features can be easily and accurately predicted once the moduli of \eqs (\ref{projectiona}) and (\ref{projectionb}) are squared, which when $\Delta\!=\!0$ yields 
\begin{eqnarray}
p_{f,x}(t)&=&\left(\frac{2\mathcal{N}J}{\Omega}\right)^2\,\!\tau^{x_0+x\!-\!2}\!\sin^2{\frac{\Omega}{2}t}\,\,,\\
p_{a,x}(t)&=&\left[\delta_{x,x_0}\!+\!\mathcal{N} \tau^{\frac{x_0\!+\!x\!-\!2}{2}}\!\left(\cos{\frac{\Omega}{2}t}\!-\!1\right)\right]^2\,\,,
\end{eqnarray}
where $p_{f,x}$ ($p_{a,x}$) stands for the photonic (atomic) probability excitation. 

It is worth mentioning that one can give a pictorial description of the above dynamics in terms of breathing-mode behaviors. While the field bound mode exhibits pure ``transverse breathing" (with respect to the array axis), the excitonic mode in addition to this also shows ``longitudinal breathing" since the atomic excitation somehow cyclically propagates along the array axis and localizes again on the starting atom. 

The setting $\eta\!=\!-0.25$ in \fig3 was chosen in order to better highlight these phenomena. Higher values of $\eta$ reduce the number of involved cavity sites [we recall that the bound mode characteristic length is given by $\lambda\!=\!1/\log[(1 \!+\! \eta)/(1 - \eta)]$, see \eq(\ref{local}) and \fig2]. For lower values of $\eta$ more cavities get involved in the dynamics but the overlap between the excitonic bound mode and the initial state shrinks. As the component of the initial state that is orthogonal to the bound mode remains frozen in the light of \eq(\ref{Heff}), such overlap clearly affects the maximum amount of energy that can be exchanged between atoms and photons. This can be seen in \fig4, where for different $\eta$'s we plot the overall atomic (photonic) excitation probability $\sum_x p_{a,x}$ ($\sum_x p_{f,x}$) against time. Remarkably, in full analogy with a standard JC model \cite{JC} the atom-field exchange of energy occurs so that the initially excited atom fully retrieves the energy released to the array regardless of the array size. Notice that this takes place at a rate given by the very same Rabi frequency associated with a single isolated cavity.
\begin{figure}
\bigskip
 \includegraphics[width=0.45\textwidth]{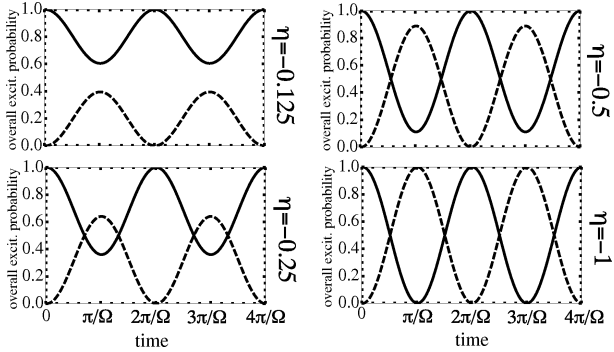}
\caption{Overall atomic (solid line) and photonic (dashed line) excitation probabilities against time for $\eta\!=\!-0.125,-0.25,-0.5,1$. The initial state is the one where atom 1 is fully excited. We have set $N\!=\!101$ and (in units of $J$) $\kappa\!=\!100$, $\omega_f\!=\!1000$, $\Delta\!=\!0$. \label{Fig5}}
\end{figure}
As anticipated, the higher $\eta$ the larger is the exchanged amount of energy. In particular, in the limit $\eta\!\rightarrow\!-1$ the field is able to absorb the entire energy initially stored in atom 1. In such a limit, indeed, the coupling between cavities 1 and 2 is fully broken. Hence, a standard JC dynamics takes place within cavity 1. This suggests that one can regard the setting presented in this work as a sort of \emph{degenerate} single-cavity JC model, which can describe the intermediate situation between a standard single-cavity JC and a uniform-hopping array.

When the initial localized excitation is purely photonic, energy exchange proceeds analogously. However, while the excitonic distribution evolves with time exactly as the photonic one in the above case (thus exhibiting pure transverse breathing) an analogous argument does not hold for the photonic distribution. This exhibits a behavior more complex than those in \fig3 and, remarkably, it spreads with time over the entire array. This can be seen with the help of \eq(\ref{Heff}): while the unbound atomic normal modes all have the same frequencies, so do not the photonic ones. Hence, unlike the process in \fig3 where the part of the initial wavefunction not overlapping the atomic local mode remains frozen, when the initial excitation is photonic the non-overlapping part freely propagates along the array in the form of photonic excitation.

\section{Conclusions} \label{conclusions}

In this work, we have considered an array of coupled cavities and atoms with interspersed hopping strengths. We have first presented analytical solutions for the normal eigenmodes of the free field Hamiltonian under open BCs to highlight the emergence of two continuous bands with a bound mode occurring at the center of their energy gap. In contrast to uniform- and parabolic-couplings arrays \cite{greentree} where in the strong-hopping regime the atomic dynamics is frozen, we have shown that depending on which cavity is initially excited a significant exchange of energy between atoms and photons can arise. The associated dynamics is basically the one occurring with a standard JC model, where the aforementioned photonic bound mode and its excitonic analogue play the roles of the cavity mode and two-level atom, respectively. Remarkably, the Rabi frequency associated with such effective JC-type dynamics is the same as the one associated with a single isolated cavity. In real space, an excitation initially localized within one cavity periodically spreads over nearby cavity sites of the same parity, in the form of both photonic and atomic excitations, and then localizes back on the starting site so that the initial conditions are retrieved. Interestingly, there is an intermediate time instant at which while the field is fully unexcited and the initially localized exciton is spread over the characteristic range of the bound mode. 

In this work we have restricted to the case of odd $N$. Indeed, in our staggered tight-binding model with open BCs the even and odd cases cannot be treated on the same footing like with cyclic conditions \cite{sun}. In the even case, while major features such as the presence of a band gap with a discrete level at its center still hold the discrete level, when present, becomes twofold  \cite{doubledeg}. We have thus focused on the odd case merely for the sake of argument in order to better highlight the physical effects that we have presented. A comprehensive treatment of both cases will be the subject of a future publication \cite{ciccarello}.

Localized (bound) normal modes often occur in solid-state physics \cite{madelung} typically in the vicinity of localized defects or impurities that break the translational invariance of the host lattice. In a similar vein, they also appear in various CCAs scenarios such as arrays with one \cite{longo} or two \cite{sun-quasi-bound} impurity atoms, T-type arrays with a single impurity atom \cite{shi} and in CCAs with one or two detuned cavities \cite{nori-no-atomi}. Here, the bound mode responsible for the phenomena that we have presented arises in a somewhat different way since no impurities or defects are present. Rather, its emergence is essentially a pure boundary effect stemming solely from the finiteness of the array length (we recall that under cyclic BCs this mode is absent \cite{sun}). As such, aside from the specific context here addressed the present work provides a paradigmatic example of boundary effects in a CCAs scenario.

Concerning an experimental test of the phenomena presented in this paper, arguments analogous to those recently discussed elsewhere for CCAs with controllable hopping strengths \cite{greentree,sun-controllable} hold here as well. It is important to point out that even though we have often assumed large-size arrays in our numerical examples  all the discussed effects in fact do not depend on the number of cavities. This makes their observation feasible even with a small-size array, a setting that is widely expected to become accessible in the imminent future.

Finally, it is worth mentioning that in the one-excitation Hilbert space of the field the bound mode in \eq(\ref{local}) in fact defines an invariant state of the free field Hamiltonian $\hat{H}_f$. In particular, notice that it makes the hopping part of the free field Hamiltonian (\ref{H0f}) effectively vanish (this is the reason why the Rabi frequency associated with the effective JC-type dynamics is the same as the one associated with a single isolated cavity). Invariant subspaces are states able to inhibit the transport of excitations through a quantum network. As such, they play a major role, although indirect, in some models recently proposed to explain the observed highly efficient excitation transfer in light-harvesting complexes \cite{caruso}. Our findings provide a novel mechanism and context (of course well different from the aforementioned ones) where invariant subspaces can also play a significant role in excitation transport. 

\begin{acknowledgments}
Fruitful discussions with  D.~Burgarth, G.~Falci, G.~M.~Palma, M.~Paternostro and M. Zarcone are gratefully acknowledged.
\end{acknowledgments}

\appendix

\section{\label{AppA}Useful formulas}

In this Appendix, we prove the three identities (to be used in the following Appendixes)
\begin{eqnarray}
\,\,\sin{(k x+\vartheta_k)}=\frac{\kappa\,(1\!-\!\eta)}{\varepsilon_k}\left\{\sin[k(x\!-\!1)]-\tau\sin(k x)\right\}\label{seno}\,\,,\,\,\,\,\,\,\,\,\,\,\,\,\,\,\,\,\,\\
\sum_{x=1}^{\frac{N\pm1}{2}}\sin (k x)\sin(k' x) \!=\! \frac{N + 1}{4}\,\delta_{k k'}\,\,,\,\,\,\,\,\,\,\,\,\,\,\,\,\,\,\,\,\,\,\,\,\,\,\,\,\,\,\,\,\,\,\,\,\,\,\,\,\,\,\,\,\,\,\,\,\,\,\,\,\,\,\,\,\,\,\,\,\,\,\,\,\label{sommeseni1}\\
\sum_{x=1}^{\frac{N+1}{2}}\left\{\sin [k (x \!- \!1)] \sin(k'x)\right.\nonumber\,\,\,\,\,\,\,\,\,\,\,\,\,\,\,\,\,\,\,\,\,\,\,\,\,\,\,\,\,\,\,\,\,\,\,\,\,\,\,\,\,\,\,\,\,\,\,\,\,\,\,\,\,\,\,\,\,\,\,\,\,\,\,\,\,\,\,\,\,\,\,\,\,\,\,\,\,\,\,\,\\
\left.\,\,\,+\sin [k' (x \!- \!1)] \sin(kx)\right\} \!=\! \frac{N \!+ \!1}{2}\cos k\,\delta_{k k'}\,,\,\,\,\,\,\,\,\,\,\,\,\,\,\label{sommeseni2}
\end{eqnarray}
where $k\!=\!2\pi m/(N\!+\!1)$ with $m\!=\!1,...,(N\!-\!1)/2$ (and similarly for $k'$ with $m'$ being the associated integer). 
\eq(\ref{seno}) is straightforwardly checked by expressing the sine function in terms of complex exponentials and then replacing $e^{\pm i \vartheta_k}$ through \eq(\ref{tetak}) as
\begin{eqnarray}
\sin{(k x+\vartheta_k)}&=&\frac{\kappa\,(1\!-\!\eta)}{\varepsilon_k}\,\,\frac{e^{i k x} (e^{-i k}\!-\!\tau)\!-\!e^{-i k x} (e^{i k}\!-\!\tau)}{2i}\,\,\,\,\,\,\,\,\,\,\,\,\,\,\,\,\,\nonumber\\
&=&\frac{\kappa\,(1\!-\!\eta)}{\varepsilon_k}\left\{\sin[k(x\!-\!1)]-\tau\sin(k x)\right\}\label{seno2}\,\,.
\end{eqnarray}
As for the remaining identities, using the well-known sum formula for geometric series it turns out that
\begin{equation}\label{geosum}
\sum_{x=1}^{\frac{N+1}{2}}e^{i k x}=\sum_{x=1}^{\frac{N+1}{2}}e^{\frac{2\pi i m x}{N+1}}=\!\left\{\begin{array}{cc}
\frac{N+1}{2}    \,\,\,\,\,\,\,\, \,\,\,\,\,\,\, \,\,\,\,  \,\,\,\,\,\,\,\,\,\,\,\,m=0     \\
          \frac{ \left[-1+(-1)^m\right] \,(1-e^{i k})}{2(1-\cos k)} \,\,\,\,\,\,\,m\neq0 \end{array}\right.\,\,.
\end{equation}
By taking the real part of \eq(\ref{geosum}) the corresponding formula for the cosine sum is obtained as
\begin{equation}\label{geosumcos}
\sum_{x=1}^{\frac{N+1}{2}}\cos{( k x)}=\!\left\{\begin{array}{cc}
\frac{N+1}{2}    \,\,\,\,\,\,\,\,\,\,\,\,\,\,\,\,m=0     \\
          \frac{-1+(-1)^m}{2} \,\,\,\,\,\,\,m\neq0 \end{array}\right.\,\,.
\end{equation}
Upon application of prosthaphaeresis formulas each product of sines appearing in the sum (\ref{sommeseni1}) can be decomposed as the sum of two cosines so as to yield that
\begin{eqnarray}
\sum_{x=1}^{\frac{N\pm1}{2}}\sin (k x)\sin(k' x) \!&=&\! \sum_{x=1}^{\frac{N\pm1}{2}}\frac{\cos[(k\!-\!k')x]\!-\!\cos[(k\!+\!k')x]}{2}\nonumber\\
\!&=&\!\left\{\begin{array}{cc}
\frac{1}{2}\! \left\{\frac{N\!+\!1}{2}\!-\!\left[\!-1\!+\!(-1)^{2m'}\right]\right\}\!=\!\frac{N\!+\!1}{4}   \,\,\,\,m=m'     \\
         0 \,\,\,\,\,\,\, \,\,\, \,\,\,\,\,\,\, \,\,\,\,\,\,\, \,\,\,\,\,\,\, \,\,\,\,\,\,\, \,\,\,\,\,\,\, \,\,\,\,\,\,\, \,\,\,\,\,\,\, \,\,\,\,\,\,\,m\neq m' \end{array}\right.\nonumber\\
         \!&=&\!\frac{N + 1}{4}\,\delta_{k k'}\,\,,\,\,\,\,\,\,\,\,\,\,\,\,\,\,\,\label{sommeseni1proof}
\end{eqnarray}
where we have used (\ref{geosumcos}) to replace the cosine sums. 

\eq(\ref{sommeseni1}) is thus proven (note that the sum remains unchanged if the upper bound $(N\!+\!1)/2$ is replaced with $(N\!-\!1)/2$ since the term corresponding to $m\!=\!(N\!+\!1)/2$ clearly vanishes).

To demonstrate (\ref{sommeseni2}), we first use \eq(\ref{geosum}) in the case $m\neq0$ in order to derive the identity
\begin{eqnarray}\label{geosumcos2}
\sum_{x=1}^{\frac{N+1}{2}}\cos{[ (k\!\pm\! k'\!) x\!-\!k]}\!&=&\!{\rm Re}\left[e^{-i k}\sum_{x=1}^{\frac{N+1}{2}}e^{i (k\pm k') x}\right]\nonumber\\
\!&=&\!\left[-1+\!(-1)^{m+m'}\right]{\rm Re}\left[\frac{1}{e^{i k}\!-\! e^{\mp i k'}}\right]\nonumber\\
\!&=&\!\frac{\left[-1+\!(-1)^{m+m'}\right]\left(\cos k \!-\!\cos k'\right)}{\left(\cos k\!-\!\cos k'\right)^2\!+\!\left(\sin k\!\pm\!\sin k'\right)^2}\,\,,\,\,\, \,\,\,\,\,\,\, \,\
\end{eqnarray}
which holds for $k\neq k'$.
For $k\!=\!k'$ we obtain that
\begin{eqnarray}\label{geosumcos3}
\sum_{x=1}^{\frac{N+1}{2}}\cos{[ (k\!\pm\! k'\!) x\!-\!k]}|_{k=k'}\!&=&\!\left\{\begin{array}{cc}
\sum_{x=1}^{\frac{N+1}{2}}\cos{[ k(2x\!-\! 1\!)]}\!=\!0   \\
       \frac{N+1}{2}\cos k \,\,\,\,\,\,\,\,\,\,\,\,\,\,\,\,\,\,\,\,\,\,\,\,\,\,\,\,\,\,\,\,\,\end{array}\,\,.\,\,\,\,\,\,\,\right.
\end{eqnarray}
In the upper case (corresponding to the case $k\!+\!k'$) the sum vanishes since clearly so does the sum over both $\sin(2 k x)$ and $\cos(2 k x)$.
Use of prosthaphaeresis formulas allows to decompose each product of sines in \eq(\ref{sommeseni2}) in terms of a cosine sum. Thereby, the first sum in the left-hand side of (\ref{sommeseni2}) can be decomposed as
\begin{eqnarray}
\!\sum_{x=1}^{\frac{N+1}{2}}\!\sin [k (x \!- \!1\!)] \!\sin k'x\!=\! \frac{\cos[(k\!-\!k')x\!-\!k]\!-\!\cos[(k\!+\!k')x\!-k]}{2},\,\,\,\,\,\,\,\label{sommeseni2prosta}
\end{eqnarray}
while the decomposition of the second sum in (\ref{sommeseni2}) is obtained from (\ref{sommeseni2prosta}) by exchanging $k$ with $k'$. It is now clear that for any $k\!\neq\! k'$ the left-hand side of (\ref{sommeseni2}) vanishes given that an exchange of $k$ with $k'$ evidently transforms (\ref{geosumcos2}) into its opposite. On the other hand, it is immediate to see that when $k\!=\!k'$ \eqs(\ref{geosumcos3}) and (\ref{sommeseni2prosta}) entail that the left-hand side of (\ref{sommeseni2}) reduces to $(N\!+\!1)/2 \cos k \delta_{k,k'}$. The proof of identity (\ref{sommeseni2}) is thus complete.

\section{\label{AppB}Orthonormality conditions}

Here, we demonstrate that the set of discrete functions specifying the normal operators in \eqs(\ref{local}) and (\ref{states}) fulfill orthonormality conditions. To this aim, we first set a suitable notation (to be used in the next appendixes as well). Firstly, we relabel each continuous-band normal mode associated with $\{k,\pm\}$ as $\{k,\pm1\}$, \ie we replace $\pm$ with the {\it numerical} index $\mu\!=\!\pm1$.
The expansion of each normal annihilation operator is rewritten as
\begin{eqnarray}
\hat{\alpha}_\ell&=&\sum_{x=1}^N \varphi_{\ell,x} \,\hat{a}_x=\sum_{x=1}^{\frac{N+1}{2}} \varphi_{\ell,2x-1} \,\hat{a}_{2x-1}+\sum_{x=1}^{\frac{N-1}{2}} \varphi_{\ell,2x} \,\hat{a}_{2x}\,\,,\label{expell}\\
\hat{\alpha}_{k\mu}&=&\sum_{x=1}^N \varphi_{k \mu,x} \,\hat{a}_x=\sum_{x=1}^{\frac{N+1}{2}} \varphi_{k \mu,2x-1} \,\hat{a}_{2x-1}+\sum_{x=1}^{\frac{N-1}{2}} \varphi_{k \mu,2x} \,\hat{a}_{2x}\,\,,\,\,\,\,\,\,\,\,\,\,\,\,\,\label{expkmu}
\end{eqnarray}
where by virtue of \eqs(\ref{local}) and (\ref{states}) the discrete real functions $\varphi$'s are defined as
\begin{eqnarray} 
\varphi_{\ell,2x-1}\!&=&\!\mathcal{A}\,\tau^{x-1}\,,\,\,\,\,\,\,\,\,\,\,\,\,\,\,\,\,\,\,\,\,\,\,\,\,\,\,\,\,\,\,\varphi_{\ell,2x}\!=\!0\,\,,\label{phi1}\\ 
\varphi_{k \mu,2x-1}\!&=&\! \mu\,\mathcal{B}\sin{(k x\!+\!\vartheta_k)}\,,\,\,\,\,\,\,\varphi_{k \mu,2x}\!=\!\mathcal{B}\,\sin{(k x)}\,\,,\,\,\,\,\label{phi2}
 \end{eqnarray}
where for compactness of notation we have set 
\begin{eqnarray}
\mathcal{A}&=&\frac{2}{\eta\!-\!1}\sqrt{\frac{\eta}{\tau^{N\!+\!1}\!-\!1}}\label{acal}\,\,,\\
\mathcal{B}&=&\sqrt{\frac{2}{N+1}}\,\,.
\end{eqnarray}
We recall that $\tau$ is defined according to \eq(\ref{tau}). Our goal is to prove that the set of $N$ discrete functions $\{{\varphi_{\ell,x}},\varphi_{k\mu,x}\}$ satisfy the orthonormality conditions
\begin{eqnarray}
\sum_{x=1}^{N}\varphi_{\ell,x}\,\varphi_{\ell,x}&=&1\,\,,\label{ort1}\\
\sum_{x=1}^{N}\varphi_{\ell,x}\,\varphi_{k\mu,x}&=&0\,\,,\label{ort2}\\
\sum_{x=1}^{N}\varphi_{k\mu,x}\,\varphi_{k'\mu',x}&=&\delta_{kk'}\delta_{\mu\mu'}\label{ort3}\,\,.
\end{eqnarray}

Upon use of \eqs(\ref{phi1}) and (\ref{acal}) along with the well-known sum formula for geometric series the following identity holds
\begin{equation}
\sum_{x=1}^{\frac{N+1}{2}}\tau^{2(x-1)}\!=\!\frac{\tau^{N+1}-1}{\tau^2-1}=\frac{(\eta\!-\!1)^2(\tau^{N\!+\!1}\!-\!1)}{4\eta}=\mathcal{A}^{-2}\,\,.
\end{equation}
Hence, it turns out that 
\begin{eqnarray}
\sum_{x=1}^{N}\varphi_{\ell,x}\,\varphi_{\ell,x}&=&\sum_{x=1}^{\frac{N+1}{2}}\varphi_{\ell,2x-1}^2=\mathcal{A}^2\sum_{x=1}^{\frac{N+1}{2}}\tau^{2(x-1)}=1\label{ort1-2}\,\,,
\end{eqnarray}
which proves \eq(\ref{ort1}). 

Using now identity (\ref{seno}) along with \eqs(\ref{phi1}) and (\ref{phi2}),  the left-hand side of \eq(\ref{ort2}) can be arranged as
\begin{eqnarray}\label{ort2-2}
\sum_{x=1}^{N}\varphi_{\ell,x}\,\varphi_{k\mu,x}&\!=\!&\mu\mathcal{A\, B}\!\sum_{x=1}^{\frac{N+1}{2}}\tau^{x\!-\!1}\sin{(k x\!+\!\vartheta_k)}\nonumber\\
&\!\propto\!&\sum_{x=1}^{\frac{N+1}{2}}\left\{\tau^{x-1}\!\sin[k(x\!-\!1)]\!-\!\tau^{x}\!\sin(k x)\right\}\nonumber\\
&\!\propto\!&\sum_{x=0}^{\frac{N-1}{2}}\!\tau^{x}\!\sin(k x)-\sum_{x=1}^{\frac{N+1}{2}}\!\tau^{x}\!\sin(k x)\!=\!0\,\,.
\end{eqnarray}
where we have used the fact that in the second sum on the last line either of the terms corresponding to $x\!=\!1$ and $x\!=\!(N\!+\!1)/2$ vanishes [we recall that $k\!=\!2\pi m/(N\!+\!1)$, see Section \ref{staggered}]. Hence, \eq(\ref{ort2}) is proven.

As for identity (\ref{ort3}), using (\ref{phi2}) the left-hand side of \eq(\ref{ort3}) can be written as
\begin{eqnarray}\label{ort3-2}
\sum_{x=1}^N \varphi_{k\mu,x}\varphi_{k'\mu',x}\!=\!\mathcal{B}^2 \left[\mu\mu'\sum_{x=1}^{\frac{N+1}{2}} \sin{(kx\!+\!\vartheta_k)}\sin{(k'x\!+\!\vartheta_{k'})}\right.\,\,\,\,\,\,\,\,\nonumber\\
\!+\!\left. \sum_{x=1}^{\frac{N-1}{2}} \sin{(kx)}\sin{(k'x)}\right]\,\,.\,\,\,\,\,\,\,
\end{eqnarray}
Upon use of \eqs(\ref{tetak}), (\ref{sommeseni1}) and (\ref{sommeseni2}), the first sum within square brackets on the right-hand side is given by
\begin{eqnarray}\label{sumsenteta}
\sum_{x=1}^{\frac{N+1}{2}} \sin{(kx\!+\!\vartheta_k)}\sin{(k'x\!+\!\vartheta_{k'})}\!&\!=\!&\!\left[\!\frac{\kappa\,(1\!-\!\eta)}{\varepsilon_k}\right]^2\!(1\!+\!\tau^2\!-\!2\tau\cos k)\!\nonumber\\
&&\times\,\frac{N\!+\!1}{4}\delta_{kk'}\!=\!\frac{N\!+\!1}{4}\delta_{kk'}\,,\,\,\,\,\,\,\,\,\,\,\,\,\,\,\,
\end{eqnarray}
where we have used the identity 
\begin{equation}\label{identitytau}
1\!+\!\tau^2\!-\!2\tau\cos k=\!\frac{\varepsilon_k^2}{\kappa^2\,(1\!-\!\eta)^2}\,\,,
\end{equation}
which can be checked straightforwardly through definitions (\ref{tau}) and (\ref{epsk}) and application of half-angle formulae.
Using \eq(\ref{sommeseni1}) along with \eq(\ref{sumsenteta}), \eq(\ref{ort3-2}) takes the form
\begin{eqnarray}\label{ort3-3}
\sum_{x=1}^N \varphi_{k\mu,x}\varphi_{k'\mu',x}\!=\!\mu\mu' \frac{\delta_{kk'}}{2}+\frac{\delta_{kk'}}{2}=\delta_{kk'}\delta_{\mu\mu'}\,\,,
\end{eqnarray}
which shows that identity (\ref{ort3}) holds. 

All of the three orthonormality identities (\ref{ort1})-(\ref{ort3}) are thus proved.

\section{\label{AppC}Normal modes in the special case ${\bf \eta=0}$}

In this Appendix, we prove that the set of normal frequencies and corresponding annihilation operators appearing in the decomposition of  $\hat{H}_{ f}$ (\ref{Hf-diag}) and defined in \eqs(\ref{local})-(\ref{states}) reduce to (\ref{omegak}) and (\ref{alphak}), respectively, in the special case $\eta\!=\!0$, as expected.

Using \eqs (\ref{epsk}) and (\ref{energies}), in the case that $\eta\!=\!0$ the normal frequencies associated with $\hat{\alpha}_{k+}$ and $\hat{\alpha}_{k-}$ become, respectively
\begin{eqnarray}
\omega_{k+}|_{\eta\!=\!0}&=&\omega_f\!+\!2\kappa\cos(k/2)\,\,,\label{en+}\\
\omega_{k-}|_{\eta\!=\!0}\!&=&\omega_f\!-\!2\kappa\cos(k/2)=\omega_f\!+\!2\kappa\cos\left(\frac{k}{2}\!-\!\pi\right)\label{en-}\,\,.
\end{eqnarray}
According to identity (\ref{seno}), proven in Appendix \ref{AppA}, for $\eta\!=\!0$ the quantity $\sin(k x\!+\!\vartheta_k)$ becomes proportional to the sum of two sines [$\tau\!\rightarrow\!-1$ according to \eq(\ref{tau})]. Upon application of the corresponding prosthaphaeresis formula we obtain that
\begin{eqnarray}\label{senoprosta}
\sin(k x\!+\!\vartheta_k)|_{\eta\!=\!0}\!&=&\!\frac{2\sin\left(\frac{kx-k+kx}{2}\right)\cos\frac{k}{2}}{2\cos\frac{k}{2}}=\sin\left[\frac{k}{2}(2x\!-\!1)\right]\nonumber\\
&=&-\sin\left[\left(\frac{k}{2}\!-\!\pi\right)(2x\!-\!1)\right]\,\,.
\end{eqnarray}
Also, $\sin(k x)$ can be arranged in either of the equivalent forms
\begin{equation}\label{senopari}
\sin(k x)=\sin\left[\frac{k}{2}(2x)\right]=\sin\left[\left(\frac{k}{2}\!-\!\pi\right)(2x)\right]\,\,.
\end{equation}
In the light of \eqs(\ref{states}), (\ref{senoprosta}) and (\ref{senopari}) $\hat{\alpha}_{k\pm}$ can be arranged as
\begin{eqnarray}
\hat{\alpha}_{k+}|_{\eta\!=\!0}\!&=&\!\!\sqrt{\frac{2}{N\!+\!1}} \sum_{x=1}^N\,\sin\left[\frac{k}{2}x\right] \,\hat{a}_{ x} \label{alphak+}\,\,,\\
\hat{\alpha}_{k-}|_{\eta\!=\!0}\!&=&\!\!\sqrt{\frac{2}{N\!+\!1}} \sum_{x=1}^N\,\sin\left[\left(\frac{k}{2}\!-\!\pi\right)x\right] \label{alphak-}\,\hat{a}_{ x} \,\,.
\end{eqnarray}
Given that $k\!=\!2\pi m/(N\!+\!1)$ with $m\!=\!1,...,(N\!-\!1)/2$ (\cf Section \ref{staggered}) the normal frequencies (\ref{en+}) and associated operators (\ref{alphak+}) coincide with the first $(N\!-\!1)/2$ operators and respective frequencies in \eqs (\ref{omegak}) and (\ref{alphak}), respectively. 

As for the normal modes specified by \eqs(\ref{en-}) and (\ref{alphak-}), one can replace the argument of the cosine and sine functions with its opposite. This is possible since while the cosine in \eq(\ref{en-}) remains unchanged the normal operator (\ref{alphak-}) is multiplied by an irrelevant global phase factor. The associated wave vectors thus become
\begin{eqnarray}
\!-\left(\frac{k}{2}\!-\!\pi\right)\!=\!\left\{\frac{\pi\,(N\!+\!1\!-\!m)}{N\!+\!1}\right\}_{m=1}^{\!\frac{N-1}{2}}\!=\!\frac{\pi N}{N\!+\!1},\!...,\frac{\pi\left[ (N\!+\!1)/2\!+\!1\right]}{N+1}.\,\,\,\,\,\,\label{wavevectors}
\end{eqnarray}
\eq(\ref{wavevectors}) shows that the normal modes corresponding to \eqs(\ref{en-}) and (\ref{alphak-}) coincide with the last $(N\!-\!1)/2$ modes specified by \eqs(\ref{omegak}) and (\ref{alphak}).

As for the bound mode (\ref{local}), for $\eta\!=\!0$ this reduces to
\begin{eqnarray}
\hat{\alpha}_{\ell}|_{\eta=0}\!&=&\lim_{\eta\to0}\left[\!\frac{2}{\eta\!-\!1}\sqrt{\frac{\eta}{\tau^{N\!+\!1}\!-\!1}}\right]\,\,\sum_{x=1}^{\frac{N+1}{2}}\,(-1)^{x-1}\,\hat{a}_{2x-1}\,\,\nonumber\\
&=&-\sqrt{\frac{2}{N\!+\!1}}\sum_{x=1}^{N}\sin\left[\frac{\pi\, [(N\!+\!1)/2]}{N\!+\!1}\,x\right]\hat{a}_x\label{local2}\,\,,
\end{eqnarray} 
thereby coinciding with (\ref{alphak}) for $m\!=\!(N\!+\!1)/2$ (up to an irrelevant global phase factor). On the other hand, for $m\!=\!(N\!+\!1)/2$ (\ref{omegak}) trivially reduces to $\omega_f$. 
Hence, in the special case $\eta\!=\!0$ all of the normal modes specified by \eqs(\ref{local})-(\ref{states}) correctly reduce to the normal modes with uniform hopping rates.
\section{\label{AppD}Free-field normal-mode decomposition}
In this Appendix, we will explicitly check the validity of \eq(\ref{Hf-diag}), \ie the normal-mode decomposition of the free field Hamiltonian $\hat{H}_f$. Due to the orthonormality conditions (\ref{ort1})-(\ref{ort3}), the (real) matrix of the coefficients that define the normal operators [\cf \eqs(\ref{phi1})-(\ref{phi2})] is orthogonal. Its inverse thus coincides with its transpose, which yields that
\begin{eqnarray}
\hat{a}_x=\varphi_{\ell,x} \,\hat{\alpha}_\ell+\sum_{k,\mu}\varphi_{k\mu,x}\,\hat{\alpha}_{k\mu}\,\,.\label{ax}
\end{eqnarray}
By using this equation to express operators $\{\hat{a}_x\}$ in terms of normal operators, the free field Hamiltonian (\ref{H0f}) becomes
\begin{widetext}
\begin{eqnarray} 
\hat{H}_f &=& \omega_f \left\{\left(\sum^N_{x=1}\varphi_{\ell x}^2\right)
       \hat{\alpha}_\ell^{\dagger}\hat{\alpha}_\ell+\sum_{k\mu}\sum_{k'\mu'} \left(\sum^N_{x=1}\varphi_{k\mu,x}\varphi_{k'\mu',x}\right)
       \hat{\alpha}_{k\mu}^{\dagger}\hat{\alpha}_{k'\mu'}+\sum_{k\mu}\left(\sum^N_{x=1}\varphi_{\ell,x}\varphi_{k\mu,x}\right)
      \left (\hat{\alpha}_{\ell}^{\dagger}\hat{\alpha}_{k\mu}+{\rm h.c.}\right)\right\}\nonumber\\
        \!& &+2  \left( \sum^{N-1}_{x=1}\, \rho_x\,\varphi_{\ell,x + 1}\varphi_{\ell,x} \right)\hat{\alpha}_\ell^{\dagger}\hat{\alpha}_\ell + \sum_{k\mu}\sum_{k'\mu'} \,\left(\sum^{N-1}_{x=1}\rho_x \, f_{k\mu,x}^{k'\!\mu'}\right) 
       \hat{\alpha}_{k\mu}^{\dagger} \hat{\alpha}_{k'\mu'} + \sum_{k\mu} \, \left(\sum^{N-1}_{x=1}\rho_x \,g_{k\mu,x}\right)
       	     \left( \hat{\alpha}_{\ell}^{\dagger}\hat{\alpha}_{k\mu} + \rm{h.c.}\right)\,\,,\,\,\,\,\,\,\,\,\,\label{hf2}
     \end{eqnarray}
\end{widetext}
where we have set 
\begin{eqnarray}
\rho_x&=&-\kappa\left[1-(-1)^x\eta\label{rho}\right]\,\,,\\
 f_{k\mu,x}^{k'\!\mu'}&=&\varphi_{k\mu,x+1}\,\varphi_{k'\mu',x}+\varphi_{k\mu,x}\,\varphi_{k'\mu',x+1}\,\,,\label{f}\\
 g_{k\mu,x}&=&\varphi_{\ell,x+1}\,\varphi_{k\mu,x}+\varphi_{\ell,x}\,\varphi_{k\mu,x+1}\,\,.\label{g}
\end{eqnarray}
The three sums over $x$ between brackets on the first line of \eq(\ref{hf2}) coincide with the left-hand sides of the orthonormality identities (\ref{ort1})-(\ref{ort3}). Moreover, due to \eq(\ref{phi1}) the coefficient of $\hat{\alpha}_\ell^{\dagger}\hat{\alpha}_\ell$ on the second line clearly vanishes. Using these facts, \eq(\ref{hf2}) considerably simplifies as
\begin{eqnarray} 
\hat{H}_f &=& \omega_f \left(
       \hat{\alpha}_\ell^{\dagger}\hat{\alpha}_\ell+\sum_{k\mu}
       \hat{\alpha}_{k\mu}^{\dagger}\hat{\alpha}_{k\mu}\right)+ \sum_{k\mu}\sum_{k'\mu'} \,\left(\sum^{N-1}_{x=1}\rho_x \, f_{k\mu,x}^{k'\!\mu'}\right) 
       \hat{\alpha}_{k\mu}^{\dagger} \hat{\alpha}_{k'\mu'} \nonumber\\
        \!& &+ \sum_{k\mu} \, \left(\sum^{N-1}_{x=1}\rho_x \,g_{k\mu,x}\right)
       	     \left( \hat{\alpha}_{\ell}^{\dagger}\hat{\alpha}_{k\mu} + \rm{h.c.}\right)\,\,.\label{hf3}
     \end{eqnarray}
A comparison between \eqs(\ref{hf3}) and (\ref{Hf-diag}) with the help of \eqs(\ref{epsk}) and (\ref{energies}) shows that the proof of (\ref{Hf-diag}) is now reduced to demonstrating the identities
\begin{eqnarray} 
\sum_{x=1}^{N-1}\rho_x\,g_{k\mu,x}&=&0\label{g1}\,\,,\\
\sum_{x=1}^{N-1}\rho_x\, f_{k\mu,x}^{k'\!\mu'}&=&-\mu\,\varepsilon_k\,\delta_{k k'}\delta_{\mu\mu'}\,\,.\,\,\label{f1}
\end{eqnarray}
In addition, we need to prove that $\varepsilon_k$ is given by \eq(\ref{epsk}).

As for the proof of (\ref{g1}), we use \eqs(\ref{phi1}) and (\ref{phi2}) to derive the following identities 
\begin{eqnarray}
 \sum_{x=1}^{N-1} (\pm 1)^x\varphi_{k\mu,x + 1}\varphi_{\ell,x} &=& \pm \sum_{x=1}^{\frac{N-1}{2}} \varphi_{k\mu,2x}\varphi_{\ell,2x - 1} = \pm\frac{\gamma}{\tau}\,\,,\label{g11}\\
\sum_{x=1}^{N-1}(\pm 1)^x\varphi_{k\mu,x}\varphi_{\ell,x + 1} &=& \pm \sum_{x=1}^{\frac{N-1}{2}} \varphi_{k\mu,2(x - 1)}\varphi_{\ell,2x - 1} =\gamma\,\,,\,\,\,\,\,\,\,\label{g12}\,\,,
\end{eqnarray}
where
\begin{equation}\label{gamma}
\gamma = {\mathcal{A B}} \sum_{x=1}^{\frac{N-1}{2}} \sin(kx)\,\tau^{x}\,\,.
\end{equation}
These identities along with \eqs(\ref{rho}) and (\ref{g}) allow to straightforwardly arrange the left-hand side of (\ref{g1}) in the form
\begin{equation}\label{g1final}
\sum_{x=1}^{N-1}\rho_x\,g_{k\mu,x}=\left(\frac{1}{\tau}\!+\!1\!+\!\frac{\eta}{\tau}\!-\!\eta\right)\gamma\!=\!\left[\frac{\eta\!+\!1}{\tau}\!-\!(\eta\!-\!1)\right]\gamma\!=\!0\,\,,
\end{equation}
where we have used definition (\ref{tau}). \eq(\ref{g1}) is thus demonstrated.

To prove \eq(\ref{f1}), we consider the following identities holding for any $x\!=\!1,...,(N\!-\!1)/2$
\begin{eqnarray}
 \rho_{2x\!-\!1} \varphi_{k\mu,2x}\varphi_{k'\mu',2x\!-\!1} \!=\!-\!\mu'\!\mathcal{B}^2\!\kappa \,\sin(k x)\,(1\!+\!\eta) \sin(k' \!x\!+\!\vartheta_{k'}\!),\,\,\,\,\,\,\,\,\,\,\,\,\,\,\,\label{f11}\\
\rho_{2x} \varphi_{k\mu,2x}\varphi_{k'\mu',2x\!+\!1} \!=\!-\!\mu'\!\mathcal{B}^2\!\kappa\,\sin(k x)\,(1\!-\!\eta)\sin[k' \!(x\!+\!1)\!+\!\vartheta_{k'}\!],\,\,\,\,\,\,\,\,\label{f12}
\end{eqnarray}
where we have used \eqs(\ref{phi2}) and (\ref{rho}).
\eqs(\ref{f11}) and (\ref{f12}) sum to
\begin{eqnarray}
 \rho_{2x\!-\!1} \varphi_{k\mu,2x}\varphi_{k'\mu',2x\!-\!1}\!+\!\rho_{2x} \varphi_{k\mu,2x}\varphi_{k'\mu',2x\!+\!1}\!=\!-\!\mu'\!\mathcal{B}^2\!\kappa\sin(k x)\mathcal{S}\label{sum1}\,,\,\,\,\,\,\,\,\,
\end{eqnarray}
where 
\begin{eqnarray}
\mathcal{S}=(1\!+\!\eta) \sin(k' \!x\!+\!\vartheta_{k'}\!)+(1\!-\!\eta) \sin[k' \!(x\!+\!1)\!+\!\vartheta_{k'}\!]\,\,.\label{esse}
\end{eqnarray}
Using now identities (\ref{seno}) and (\ref{identitytau}) along with definition (\ref{tau}), $\mathcal{S}$ can be arranged in the form
\begin{eqnarray}
\mathcal{S}=\frac{\kappa\,(1\!-\!\eta)}{\varepsilon_{k'}}\left\{\left[(1\!-\!\eta)\!-\!\tau(1\!+\!\eta)\right]\sin(k'x)\right.\nonumber\,\,\,\,\,\,\,\,\,\,\,\,\,\,\,\,\,\,\,\,\,\,\,\,\,\,\,\,\,\,\,\,\,\,\,\,\,\,\,\,\,\\
\left.\,\,\,\,\,\,\,\,\,\,\,\,\,\,\,\,+(\eta\!+\!1)\left(\sin[k'(x\!-\!1)]\!+\!\sin[k'(x\!+\!1)]\right)\right\}\nonumber\\
\!=\!\frac{\kappa\,(1\!-\!\eta)^2}{\varepsilon_{k'}}\left[(1\!-\!\tau^2\!-\!2\tau\cos k')\sin (k'\!x)\right]\!=\!\frac{\varepsilon_k^2}{\kappa\varepsilon_{k'}}\sin (k'\!x).\,\,\,\,\,\,\label{esse2}
\end{eqnarray}
In deriving \eq(\ref{esse2}) we have made use of prosthaphaeresis formulas to replace the sum of sines on the second line of (\ref{esse2}) with the product $2\!\cos k'\!\sin(k' x)$. Using this result, upon sum of (\ref{sum1}) over $x\!=\!1,...,(N\!-\!1)/2$ we thus obtain that
\begin{eqnarray}
 \sum_{x=1}^{\frac{N\!-\!1}{2}}\rho_{2x\!-\!1} \varphi_{k\mu,2x}\varphi_{k'\mu',2x\!-\!1}\!+\!\rho_{2x} \varphi_{k\mu,2x}\varphi_{k'\mu',2x\!+\!1}\!=\!-\frac{\mu'\varepsilon_k\,\delta_{kk'}} {2}\!\label{sum2}\,\,,\,\,\,\,\,\,\,
\end{eqnarray}
where we have used identity (\ref{sommeseni1}) to carry out the sum of $\sin(kx)\sin(k'x)$.

Consider now the left-hand side of \eq(\ref{f1}), where the sum can be split into the sums over even and odd terms. A comparison of this with the left-hand side of \eq(\ref{sum2}) should make clear that
(\ref{f1}) can be obtained by adding (\ref{sum2}) to the quantity obtained from (\ref{sum2}) through exchange of $\{k,\mu\}$ with $\{k',\mu'\}$. This argument leads to 
\begin{equation}
\sum_{x=1}^{N-1}\rho_x\, f_{k\mu,x}^{k'\!\mu'}=-\frac{(\mu\!+\!\mu')\,\varepsilon_k\,\delta_{kk'}}{2}=-\mu\,\varepsilon_k\,\delta_{kk'}\delta_{\mu\mu'}\,\,,
\end{equation}
which proves that \eq(\ref{f1}) holds.

Although we have proven both the identities (\ref{g1}) and (\ref{f1}), to finalize the explicit demonstration of \eq(\ref{Hf-diag}) we need to prove that $\varepsilon_k$ is given by \eq(\ref{epsk}) (note that in this Appendix up to this stage we have left $\varepsilon_k$ unspecified).
To accomplish this task, we point out that \eq(\ref{tetak}) entails the constraint
\begin{equation}\label{constraint}
1=\frac{\kappa^2(1-\eta)^2}{\varepsilon_k^2}\,|e^{-ik}-\tau|^2\,\,,
\end{equation}
\ie the squared modulus of the complex exponential on the left-hand side must be unitary. By bringing $\varepsilon_k^2$ to the left-hand side and calculating the squared modulus on the right-hand side \eq(\ref{constraint}) takes the form
\begin{eqnarray}\label{constraint2}
\varepsilon_k^2
&=&\kappa^2\left\{\left[(\eta\!-\!1)\cos k\!-(\eta\!+\!1)\right]^2\!+\!(\eta\!-\!1)^2\sin^2 k\right\}\nonumber\\
&=&4\kappa^2  \left(\cos^2 \frac{k}{2}\!+\!\eta^2\sin^2{\frac{k}{2}}\right)\,\,,
\end{eqnarray}
where in the last step we have made use of half-angle formulae. By taking the square root of (\ref{constraint2}) \eq(\ref{epsk}) is retrieved.
The proof of decomposition (\ref{Hf-diag}) is therefore complete.

\section{\label{AppE}Full Hamiltonian normal-mode decomposition}
Here, we prove that the full Hamiltonian (\ref{H}) can be decomposed in terms of normal operators according to \eq(\ref{HNMD}).
Having proven in Appendix \ref{AppD} that (\ref{Hf-diag}) holds, it suffices to demonstrate that
\begin{eqnarray}
\sum^N_{x=1}
      \hat{b}_x^\dagger \hat{b}_x&\!=\!&\hat{\beta}_{\ell}^{\dagger}  \hat{\beta}_{\ell}+\!\sum_{k\mu}\hat{\beta}_{k\mu}^{\dagger}  \hat{\beta}_{k\mu}\,\,,\,\,\,\,\,\,\,\,\,\,\,\,\,\,\,\,\,\,\,\,\,\,\,\,\,\,\,\,\,\,\,\,\,\,\,\,\,\,\,\,\,\,\,\,\,\,\,\,\,\,\,\label{hnmd1}\\
\sum_{x=1}^N \left(\hat{a}_x^\dagger \hat{b}_x+\hat{b}_x^\dagger \hat{a}_x\right)&\!=\!&\left( \hat{\alpha}^{\dagger}_\ell \hat{\beta}_{\ell}\!+\! \hat{\beta}_\ell^{\dagger} \hat{\alpha}_{\ell}\right)+\!\sum_{k\mu}\left( \hat{\alpha}^{\dagger}_{k\mu} \hat{\beta}_{k\mu}\!+\! \hat{\beta}_{k\mu}^{\dagger} \hat{\alpha}_{k\mu} \right).\,\,\,\,\,\,\,\,\,\,\,\label{hnmd2}
\end{eqnarray}
We first note that due to \eqs(\ref{localb}) and (\ref{statesb}) the orthonormality conditions (\ref{ort1})-(\ref{ort3}) entail that \eq(\ref{ax}) holds for the atomic operators as well, \ie
\begin{eqnarray}
\hat{b}_x=\varphi_{\ell,x} \,\hat{\beta}_\ell+\sum_{k,\mu}\varphi_{k\mu,x}\,\hat{\beta}_{k\mu}\,\,.\label{bx}
\end{eqnarray}
Using this equation to express each atomic site operator in \eq(\ref{hnmd1}) in terms of normal operators along with the orthornormality identities (\ref{ort1})-(\ref{ort3}), the proof of \eq(\ref{hnmd1}) is in fact identical to the one carried out in the case of field operators [\cf first line of \eq(\ref{hf2}) and the quantity between the first brackets on the right-hand side of \eq(\ref{hf3})].
Likewise, use of \eqs (\ref{ort1})-(\ref{ort3}) and (\ref{bx}) yields that
\begin{eqnarray}
\sum_{x=1}^N \hat{a}_x^\dagger \hat{b}_x\!&=&\!\sum_{x=1}^N\,\!\left(\varphi_{\ell,x}\hat{\alpha}^\dagger_{\ell}\!+\!\sum_{k\mu}\varphi_{k\mu,x}\hat{\alpha}^\dagger_{k\mu}\right)\left(\varphi_{\ell,x}\hat{\beta}_{\ell}\!+\!\sum_{k'\mu'}\varphi_{k'\mu',x}\,\hat{\beta}_{k\mu}\right)\nonumber\\
\!&=&\!  \left(\sum_{x=1}^N\varphi_{\ell,x}^2 \right)\hat{\alpha}^\dagger_{\ell}\hat{\beta}_\ell+\sum_{k\mu}\sum_{k'\mu'}\left(\sum_{x=1}^N\varphi_{k\mu,x}\varphi_{k'\mu',x} \right)\hat{\alpha}^\dagger_{k\mu}\hat{\beta}_{k'\mu'}\nonumber\\
\!&&+\!\sum_{k\mu}\left(\sum_{x=1}^N\varphi_{\ell,x}\varphi_{k\mu,x}\right)\left(\hat{\alpha}_\ell^\dagger\hat{\beta}_{k\mu}\!+\!\hat{\alpha}_{k\mu}^\dagger\hat{\beta}_\ell\right)\nonumber\\
\!&=&\!\hat{\alpha}^{\dagger}_\ell \hat{\beta}_{\ell}\!+\!\sum_{k\mu} \hat{\alpha}^{\dagger}_{k\mu} \hat{\beta}_{k\mu}\,\,.\label{ab}
\end{eqnarray}
By summing \eq(\ref{ab}) to its adjoint \eq(\ref{hnmd2}) is retrieved. The proof of normal-mode decomposition (\ref{HNMD}) is thus complete. Note that the uniformity of the atomic frequencies and atom-photon coupling strengths throughout the cavity array is crucial for \eqs(\ref{hnmd1}) and (\ref{hnmd2}), and hence (\ref{HNMD}), to hold.

\end{document}